\documentstyle[preprint,revtex]{aps}

\begin{document}

\textheight232mm
\textwidth160mm
\oddsidemargin10mm
\baselineskip5.5mm

\draft

\begin{title}
One-step model of photoemission for non-local potentials
\end{title}

\author{M. Potthoff$^{\rm a}$\cite{ea}, J. Lachnitt$^{\rm b}$,
        W. Nolting$^{\rm a}$ and J. Braun$^{\rm b}$}

\begin{instit}
$^{\rm a}$ Humboldt-Universit\"at zu Berlin, Institut f\"ur Physik,  
Berlin, Germany
\end{instit}

\begin{instit}
$^{\rm b}$ Universit\"at Osnabr\"uck, Fachbereich Physik,  
Osnabr\"uck, Germany
\end{instit}

\begin{abstract}
\baselineskip6mm
The one-step model of valence-band photoemission and inverse 
photoemission from single-crystal surfaces is reformulated for
generalized (non-local, complex and energy-dependent) potentials.
Thereby, it becomes possible to account for self-energy corrections 
taken from many-body electronic-structure calculations. The original
formulation due to Pendry and co-workers employs the KKR 
multiple-scattering theory for the calculation of the initial state. 
This prevents a straightforward generalization of the one-step model
to non-local potentials. We therefore consider the Dyson equation 
which is set up within a muffin-tin-orbitals representation as an 
alternative to obtain the initial-state Green function.
This approach requires a revision of the transition-matrix 
elements which is carried out in detail. The 
final state is considered as a time-reversed LEED state as usual. 
The proposed generalization of the one-step model allows to 
distinguish between the bare photocurrent reflecting the 
(quasi-particle) band structure and the secondary effects 
due to the (dipole) selection rules and due to the wave-vector 
and energy dependence of the transition-matrix elements.
\end{abstract}


\section{INTRODUCTION}
\label{sec:intro}

Much information on the electronic structure at crystal surfaces
is gained by ultraviolet photoemission spectroscopy (PES) 
\cite{FFW78,CL78,Ing82,CH84,Kev92} and inverse photoemission 
spectroscopy (IPE) \cite{Dos83,Dos85,BT88,Smi88,Don94}. Energy- 
and angle-resolved PES and IPE are useful tools for analyzing the 
dispersion of (quasi-particle) bands. For the sometimes rather 
involved interpretation of measured spectra, a comparison with 
theoretical results for the electronic structure is inevitable. 
However, measured energy- and angle-resolved spectra can hardly be 
compared with the calculated band structure directly: Secondary 
effects due to the wave-vector and energy dependence of the 
transition-matrix elements and due to selection rules considerably 
distort the ``bare'' spectra. Even if the primary interest rests 
on the bare quantities, such as the density of states and the 
dispersion of bands, a theory of photoemission is needed to 
build the bridge between the photoemission experiments and the 
electronic-structure calculation.

The so-called one-step model of photoemission has quite 
successfully 
fulfilled this demand in recent years. The original method as 
proposed 
by Pendry and co-workers \cite{Pen74,Pen76,HPT80} has been the 
starting point for several generalizations and improvements, which
became necessary as the experimental technique was more and more
refined and growing interest was spent on more complex materials.
Inverse photoemission in connection with a realistic model for the 
surface barrier \cite{Pen80,MR80,BT88} has enlarged the 
applicability of the one-step model. The relativistic 
generalization 
\cite{GDG83,TB84,BTB85,BTB87,HTG+93} made possible investigations 
of spin-polarized photoemission from non-magnetic materials 
\cite{TPF87,SVH88}.
The extension of the one-step model to magnetic materials resulted 
in theoretical studies of the magnetic circular and linear 
dichroism \cite{SHHF94,HHSF94}.
Photoemission from covalently bonded systems, adsorbate-covered
surfaces and from materials with a relatively open crystal 
structure brought the need
to overcome the muffin-tin potential model on which the original 
one-step formulation was based. The generalization
of photoemission theory with respect to space-filling potentials 
\cite{GBB93,GBB94a} and the development of a relativistic, 
full-potential one-step model for materials with several atoms per
unit cell \cite{GBB94b} has more or less completed recent work on
that matter.

The one-step model of (inverse) photoemission starts from a given
electronic potential which is needed for the necessary 
construction of multiple-scattering states and propagators. This
potential is taken from a separate electronic-structure calculation
and must be considered as the decisive input for the one-step
model: The peak positions in the calculated (inverse) photoemission 
spectrum directly reflect the energetics of the electronic structure 
that evolves from the potential given. The reliability of the PES 
(IPE) theory is thus intimately related to the ``quality'' of the 
potential for the system at hand.

The one-step model has been originally formulated for a 
{\em local} potential $V({\bf r})$. A local potential is provided 
by density-functional theory (DFT) within the local-density 
approximation (LDA) \cite{HK64,KS65,SK66,GL76,HL71,JG89}. In 
practice, such LDA inputs for photoemission calculations have 
been proven to yield rather satisfactory results when compared 
with experimental spectra from several materials 
\cite{SK95,Bar92,Bra96}. The 
use of LDA potentials has thus become a well-known and widely 
used concept in the photoemission theory.

On the other hand, when applying the theory to strongly correlated 
electron systems, it surely becomes doubtful whether this concept 
is still adequate since implicitly the LDA eigenenergies are
interpreted as to be the one-particle excitation energies of the 
system. It is well known that there are two possible sources of error
connected with that interpretation: Firstly, the LDA only provides
an approximate expression for the (local) exchange-correlation
potential. Secondly, even with the exact exchange-correlation
potential at hand, one is left with the problem that there is 
no known correspondence between the Kohn-Sham eigenenergies and 
the one-particle excitation energies \cite{JG89,KV83,AvB85,Bor85}.

For an in principle correct description of the excitation energies
the non-local self-energy has to be considered. This, however, 
constitutes a many-body problem. Therefore, DFT-LDA 
calculations must be supplemented by many-body
methods to arrive at a realistic description of the one-particle 
excitations in strongly correlated systems. To give an
example, let us mention the GW approximation \cite{Hed65} which 
is well suited for the case of insulators and semi-conductors 
but has also been applied successfully to transition metals 
\cite{Hed65,Ary92,RKP93,AG95}. Another approach is to consider 
Hubbard-type models where those Coulomb-interaction terms are 
included explicitly that are assumed to be treated insufficiently 
within DFT-LDA. As some recent studies of this type show, significant
improvement upon the LDA predictions for transition metals is
possible indeed 
\cite{NBDF89,SOH90,BN90,BBN92,SAS92,UIF94,NVF95,VN96}.

The self-consistent potential $U({\bf r},{\bf r}',E)$ resulting from
the GW approach, consists of a local part $V({\bf r})$ which can be 
identified with the LDA potential and of the non-local self-energy
$\Sigma({\bf r},{\bf r}',E)$ which is energy-dependent and complex. 
A tight-binding one-particle basis is used by those approaches that
start from (multi-band) Hubbard-type models. Consequently, 
the resulting generalized potential is given in the form
$U_{ii'}^{LL'}(E)$ where $i$ refers to sites, and $L$ is an 
orbital quantum number. Again, $U$ can be decomposed 
into a term $V_{ii'}^{LL'}$ resulting from the local one-particle 
potential in the Hamiltonian, and into the self-energy
$\Sigma_{ii'}^{LL'}(E)$ resulting from the interactions. Note
that transformation to the real-space representation yields a
non-local self-energy $\Sigma({\bf r},{\bf r}',E)$,
even if it is local in the tight-binding basis,
$\Sigma_{ii'}^{LL'}(E) \sim \delta_{ii'}$.

As concerns a theory of PES and IPE, we have to face the following 
problem: Strong electron correlations imply the need for self-energy
corrections to obtain a realistic description of the elementary 
excitations. Consequently, a generalized potential should be 
considered to be the starting point for the actual (inverse) 
photoemission theory. The non-locality of the potential, however, 
causes difficulties within the original formulation of the 
one-step model: The calculation \cite{Pen76} of the 
one-electron initial-state Green function $G({\bf r},{\bf r}',E)$, 
which is directly related to the (bare) photocurrent as well as the
subsequent evaluation of the matrix elements is based on the
Korringa-Kohn-Rostocker (KKR) multiple-scattering approach
\cite{Kor47,KR54} and is thus intrinsically limited to the case 
of local potentials. This seems to prevent a straightforward 
generalization of the theory to non-local potentials and thereby 
an application to strongly correlated systems.

The purpose of the present paper is to demonstrate that it 
is possible to overcome the mentioned restriction. Starting 
from Pendry's formula for the photocurrent \cite{Pen76}, we 
propose a reformulation of the one-step model such that a 
general (non-local, complex, energy-dependent) potential 
$U({\bf r},{\bf r}',E)$ can be included via the Dyson equation 
in the calculation of the initial-state Green function. For the 
final state the usual (layer-KKR) multiple-scattering theory 
\cite{Kor47,KR54,Kam67,Kam68,Pen74} is retained. Our approach 
requires a complete revision of the transition-matrix elements. 
This is carried out in detail. Compared with earlier theoretical 
approaches to photoemission 
\cite{Kub57,Ada64,Kel65,Mah70,SA71,CLRRSJ73,FE74}, 
our ansatz keeps the basic structure of the one-step model, which 
is highly desirable for means of numerical evaluation \cite{Pen76}.

In a preceding study \cite{BBN92} concerning PES and IPE from Ni 
surfaces, a first pragmatic way to include (non-local) self-energy 
corrections in the one-step model was presented. That approach, 
however, was based on some simplifying assumptions by which the 
emission from different subbands could be treated independently 
of each other. This has restricted the range of applicability to 
special regions in ${\bf k}$-space. Furthermore, the method rests 
on some presuppositions for the pole structure of the self-energy 
which were shown to be adequate for the case of Ni, but may be 
too special in other cases. With the present work we also try to 
overcome these restrictions and aim at an improved concept that 
allows to include arbitrary self-energy corrections in the 
one-step model.

The paper is organized as follows: The next section introduces
Pendry's basic formula for the photocurrent. The discussion of 
the Green function for the initial state in Sec.\ \ref{sec:igr} 
elucidates the difficulties that arise within the original formalism
for a non-local potential. Sec.\ \ref{sec:evalp} prepares for the
proposed alternative evaluation of Pendry's formula which is carried
out in detail in Sec.\ \ref{sec:ini} for the initial states, in 
Sec.\ \ref{sec:fin} for the final state, and eventually in Sec.\ 
\ref{sec:tme} where the transition-matrix elements are considered 
and all partial results are put together. Finally, Sec.\ 
\ref{sec:con} concludes our considerations.

\section{ONE-STEP MODEL OF PHOTOEMISSION}
\label{sec:osm}

The starting point for the one-step model of photoemission is 
Pendry's formula \cite{Pen76,Bor85}:

\begin{equation}
  I^{\rm PES} \propto f_{\rm F}(E_1)
  \mbox{Im} 
  \langle \epsilon_f {\bf k}_{||} \sigma | 
  G_{2\sigma}^+ \Delta G_{1\sigma}^+ \Delta^\dagger G^-_{2\sigma} | 
  \epsilon_f {\bf k}_{||} \sigma \rangle \: .
\label{eq:pendry}
\end{equation}
All relevant information on the electronic structure around the 
Fermi energy and on the one-particle excitations is included in the
``low-energy'' propagator $G_{1\sigma}^+$, i.~e.\ in the operator
representation of the one-electron retarded Green function for the
initial state:

\begin{equation}
G_{1\sigma}^+ =
\int \int | {\bf r} \sigma \rangle \: 
G_{\sigma}^+({\bf r},{\bf r}',E_1) \:
\langle {\bf r}' \sigma | \;
d{\bf r} d{\bf r}' \: .
\end{equation}
$\sigma=\uparrow, \downarrow$ refers to the z-component of electron 
spin. $E_1$ is the initial-state energy, i.~e.\ $E_1=\epsilon_f - 
\mu - \hbar \omega$, where $\epsilon_f$ denotes the one-particle 
energy of the outgoing photoelectron, $\mu$ stands for the chemical 
potential, and $\hbar \omega$ is the photon energy. The low-energy
propagator $G_{1\sigma}^+$ is directly related to the ``bare''
photocurrent and thereby represents the central physical quantity
within the one-step model.

Referring to an energy-, angle- and spin-resolved 
photoemission experiment, the state of the photoelectron at the 
detector is written as $|\epsilon_f {\bf k}_{||} \sigma \rangle$, 
where ${\bf k}_{||}$ is the component of the wave vector parallel 
to the surface. By means of the advanced Green function 
$G_{2\sigma}^-$ in operator representation and taken at the 
final-state energy $E_2 \equiv \epsilon_f$, we have 
$|f\rangle = G^-_{2\sigma} |\epsilon_f {\bf k}_{||} \sigma \rangle$ 
for the final state in Eq.\ (\ref{eq:pendry}).
Furthermore, $\Delta = (e/m_e) {\bf A}_0 \cdot {\bf p}$ is the
dipole operator, i.~e.\ the
perturbation which mediates the coupling between the initial and
the final state. $\Delta$ has been taken in the electric dipole
approximation which is well justified in the visible and ultraviolet
spectral range. ${\bf A}_0$ is the spatially constant vector 
potential inside the crystal which can be determined from classical 
macroscopic dielectric theory.

PES and IPE are complemental spectroscopies. At zero temperature 
$T=0$ photoemission probes the occupied and inverse photoemission 
the unoccupied part of the (quasi-particle) band structure. For 
PES this fact is accounted for by the Fermi function in Eq.\ 
(\ref{eq:pendry}): $f_{\rm F}(E) = 1/(\exp (E/k_BT) + 1)$. Apart from
the Fermi function the ratio $I^{\rm PES}/I^{\rm IPE}$ is known 
to depend on kinematic factors only. It is given by the energies 
and emission angles of the emitted photoelectrons and photons, 
respectively \cite{Pen80}. We can thus concentrate on PES in the
following; IPE may always be treated analogously without any 
difficulties.

The basic equation (\ref{eq:pendry}) for the one-step model of 
photoemission contains some approximations that should be discussed
briefly: First we note that the formula is eventually based on 
Fermi's golden rule from first-order time-dependent perturbation
theory with respect to $\Delta$ \cite{Bor85}. This implies that 
it yields the elastic part of the photocurrent only. So-called 
vertex renormalizations and three-particle correlation effects 
are neglected which means the exclusion of inelastic energy losses 
and corresponding quantum-mechanical interference terms (cf.\ Refs.\
\cite{Pen76,CLRRSJ73,Bor85}).

The ``high-energy'' propagator $G_{2\sigma}^-$ is understood to be 
calculated for a local potential, i.\ e.\ assuming  
$\Sigma_\sigma({\bf r},{\bf r}',E_2) = \delta({\bf r}-{\bf r}') 
\Sigma_\sigma({\bf r}, E_2)$ for the self-energy {\em at the final 
state energy $E=E_2$}. This allows us to consider the final state 
$G_{2\sigma}^-|\epsilon_f {\bf k}_{\|} \sigma \rangle$ as a 
(time-reversed) LEED state as in the conventional one-step model. 
It can thus be calculated by the standard Korringa-Kohn-Rostocker 
technique (layer-KKR) \cite{Pen74} (see Sec.\ \ref{sec:fin}). 
Furthermore, the interaction of the outgoing photoelectron with
the rest system has been neglected (sudden approximation). This
approximation is believed to hold well for not too small photon
energies. Let us emphasize that the sudden approximation for the
final state is consistent with a local self-energy at $E=E_2$.
We assume that many-body effects in the final state $G_{2\sigma}^-
| \epsilon_f {\bf k}_{\|} \sigma \rangle$ can be sufficiently well
accounted for by the inclusion of a spatially constant, complex
self-energy correction $\Sigma_f(E_2)$ in the LEED calculation as 
usual. This weakly energy-dependent self-energy correction has to 
be distinguished clearly from the (imaginary) optical potential, 
which phenomenologically takes into account inelastic corrections 
to the elastic photocurrent \cite{Bor85}, and from the (real) inner 
potential, which serves as a reference energy inside the solid with 
respect to the vacuum level \cite{HPM+95}. The inner potential, the 
optical potential and the self-energy correction for the final state 
can formally be included in the definition of a generalized 
energy-dependent inner potential:
$V_{0,2}=V_0(E_2)=V_{0{\rm r}}(E_2)+iV_{0{\rm i}}(E_2)$.

While typical values for the final-state energy $E_2$ may be several 
tens of eV or even more above the Fermi energy, the energy of the
initial state $E_1$ lies some eV below the Fermi energy. 
It is this low-energy range that our primary interest should 
concern since excitations that involve the states in the vicinity
of the Fermi energy can be of 
particular significance for typical electron-correlation effects.
These excitations are described by means of the 
Green function for the initial state $G_{1\sigma}^+$.

\section{INITIAL-STATE GREEN FUNCTION}
\label{sec:igr}

Within the conventional one-step model of Pendry and co-workers
\cite{Pen74,Pen76,HPT80} the initial-state Green function 
$G_{1\sigma}^+$ is determined for electrons moving in an 
(effective) one-particle potential as it is provided, for 
example, by DFT-LDA:

\begin{equation}
  V_{\rm LDA}({\bf r},\sigma) =
  V_{\rm e}({\bf r}) + V_{\rm H}({\bf r}) + 
  V_{\rm xc}({\bf r},\sigma) \: .
\end{equation}
Within the DFT ground-state formalism the external core potential
$V_{\rm e}$, the Hartree contribution $V_{\rm H}$ as well as the
exchange-correlation potential $V_{\rm xc}$ are local functions.
On the other hand, it is well known that for an in principle exact
description of the one-particle excitations one has to consider the
Dyson equation for the Green function \cite{JG89,Eco90}. This 
includes the non-local, complex and energy-dependent self-energy. 
The development of reasonable approximations for the self-energy 
constitutes a fairly complex many-body problem which falls outside 
the scope of this paper. Just as the LDA potential, the self-energy 
must be assumed to be a given quantity for the photoemission theory. 

Several techniques that have been developed in the past to account
for electron-correlation effects beyond DFT-LDA yield explicit 
analytical expressions for the self-energy or at least a numerical 
data set resulting from a self-consistent calculation. Either the
self-energy is given in its real-space representation, 
$\Sigma_{\sigma}({\bf r},{\bf r}',E)$, like in the GW approximation, 
for example \cite{Hed65,Ary92,RKP93,AG95}, or in a tight-binding
representation, $\Sigma_{ii'\sigma}^{LL'}(E)$, as it is frequently 
employed within many-body approaches that consider a 
degenerate-band Hubbard model with realistic Coulomb-interaction 
parameters \cite{NBDF89,SOH90,BN90,BBN92,SAS92,UIF94,NVF95,VN96}.
The many-body calculation should explicitly take into account the
presence of the solid surface: For PES and IPE the information depth 
is mainly determined by the electron inelastic mean free path in the 
solid. Due to the small attenuation length for low-energy electrons 
\cite{pow88}, PES and IPE are sensitive to a few layers from the 
surface only. However, all hitherto existing approaches concerning 
the multi-band Hubbard model refer to an infinitely extended, 
periodic crystal lattice and benefit from simplifications due to 
translational symmetry. For the single-band Hubbard model there are 
recent attempts to take into account the breakdown of translational 
symmetry at the surface \cite{PN95,PN96,PN97c}. An extension of such 
real-space many-body techniques to the multi-band case is
straightforward. 

For the following let us assume that the LDA potential $V_{\rm LDA}$
as well as the self-energy $\Sigma$ fully account for the presence 
of the surface. The local LDA potential defines the LDA Hamiltonian:

\begin{equation}
  h_{\rm LDA}({\bf r},\sigma) = \frac{{\bf p}^2}{2m_e} +
  V_{\rm LDA}({\bf r},\sigma) \: .
\end{equation}
The self-energy is assumed to be defined with respect to 
$h_{\rm LDA}$ and given in real-space representation for 
the moment. Therewith, the generalized potential that is 
obtained by adding the self-energy to the LDA potential,

\begin{equation}
  U_\sigma({\bf r},{\bf r}',E) =
  \delta ({\bf r}-{\bf r}') \: V_{\rm LDA}({\bf r},\sigma)
  + \Sigma_{\sigma}({\bf r},{\bf r}',E) \: ,
\end{equation}
in principle yields the correct one-particle excitation spectrum.
Now the initial-state Green function may be obtained as the solution
of the Dyson equation which in real-space representation reads:

\begin{equation}
  \left[ E_1 - 
  h_{\rm LDA}({\bf r},\sigma) \right]
  G^+_{\sigma}({\bf r},{\bf r}',E_1) -
  \int \Sigma_{\sigma}({\bf r},{\bf r}'',E_1) 
  G^+_{\sigma}({\bf r''},{\bf r}',E_1) d {\bf r}'' =
  \hbar \delta({\bf r}-{\bf r}') \: .
  \label{eq:eomgr}
\end{equation}

Within the original framework of the one-step model ($\Sigma 
\equiv 0$) the calculation proceeds by interpreting the Green 
function as a propagator summing over all scattering paths that 
take the electron from ${\bf r}'$ to ${\bf r}$. Applying the KKR 
multiple-scattering theory \cite{Kor47,KR54,Kam67,Kam68,Pen74} 
solves the problem to evaluate the basic formula for the photocurrent
(\ref{eq:pendry}) very efficiently and has been adopted in 
the original work \cite{Pen76} as well as (with the necessary
modifications) in all subsequent generalizations of the one-step
model \cite{GDG83,TB84,BTB85,BTB87,HTG+93,GBB93,GBB94a,GBB94b}.

A key point of this technique concerns the following expression
for the Green function in the presence of a {\em single} atom with 
spherically symmetric potential known from scattering theory 
\cite{Mes79b,Eco90,BGZ92}:

\begin{equation}
  G^{(0)}({\bf r},{\bf r}',E_1) = -i \sqrt{E_1} \sum_{lm}
  \psi_l(r_<) \psi_l^+(r_>) 
  Y_{lm}(\hat{\bf r}') Y^\ast_{lm}(\hat{\bf r}) \: .
\label{eq:grgrk}
\end{equation}
Here $r_<$, $r_>$ stands for the lesser or greater of $|{\bf r}|$
and $|{\bf r}'|$, $\psi_l(r)$ ($\psi_l^+(r)$) is the regular 
(irregular) solution of the radial Schr\"odinger equation at energy 
$E_1$, and $Y_{lm}(\hat{\bf r})$ are the spherical harmonics. 
The advantage to use the expansion (\ref{eq:grgrk}) in the basic 
equation (\ref{eq:pendry}) is obvious: One avoids to solve the
Dyson equation in the form (\ref{eq:eomgr}), and the problem 
separates with respect to the variables ${\bf r}$ and 
${\bf r}'$. The representation (\ref{eq:grgrk}) is not restricted
to a spherically symmetric potential but can be generalized for 
potentials of arbitrary shape as has been shown by Butler et al.\
\cite{BGZ92}. 

However, the proof in (the appendix of) Ref.\ \cite{BGZ92} also 
shows that a further generalization of (\ref{eq:grgrk}) to the 
case of non-local potentials seems to be impossible. Firstly, one
surely has to consider $\psi_l(r)$ and $\psi_l^+(r)$ as the (radial)
solutions of Schr\"odinger's equation for a non-local (atomic) 
potential. But even with the correct solutions, the expression 
(\ref{eq:grgrk}) does not solve the problem: The reason is that 
due to the non-locality of the self-energy the cases $r<r'$ and 
$r>r'$ are mixed and cannot be treated independently in Eq.\ 
(\ref{eq:eomgr}) to show that 
$[E_1 - h_{\rm LDA} - \Sigma(E_1)] \: G^{(0)}(E_1) = 0$ 
for $r\ne r'$. On the other hand, however, it is necessary to 
work with $r_<$ and $r_>$ in (\ref{eq:grgrk}) in order to get 
the delta function on the right-hand side of (\ref{eq:eomgr})
(cf.\ Ref.\ \cite{BGZ92}). Furthermore, one has to consider 
multiple-scattering corrections to Eq.\ (\ref{eq:grgrk}) when 
embedding the atom in the crystal lattice. However, there is no 
generalization of the KKR multiple-scattering formalism to our 
knowledge that works for a non-local lattice-periodic potential.

Within the original one-step formalism the representation 
(\ref{eq:grgrk}) is employed from the very beginning.
Consequently, for the case of non-local potentials we have
to disregard (\ref{eq:grgrk}) as well as the
multiple-scattering formalism for the calculation of the
initial-state Green function $G_{1\sigma}^+$ in Pendry's formula.

\section{ALTERNATIVE EVALUATION OF PENDRY'S FORMULA}
\label{sec:evalp}

The alternative is to solve the Dyson equation (\ref{eq:eomgr})
directly. This, however, requires a revision of the formalism to 
a large extent. Only for the calculation of the final state the 
multiple-scattering approach can be retained. In the following 
it is shown that a consequent reformulation in the sense mentioned
is possible indeed.

The Dyson equation is most conveniently solved within a 
tight-binding representation. For this purpose we choose a set 
of one-particle wave functions,

\begin{equation}
   \Phi_{iL\sigma}({\bf r}-{\bf R}_i) \equiv
   \langle {\bf r} | i L \sigma \rangle \: ,
\label{eq:basis}
\end{equation}
which span the subspace ${\cal H}_S$ that includes all s-, p-, d- 
(and f-)like eigenstates of the LDA Hamiltonian $h_{\rm LDA}$
in the vicinity (several eV) of the Fermi energy (minimal basis set):

\begin{equation}
  \sum_{ii'LL'} \Phi_{iL\sigma}({\bf r}-{\bf R}_{i})
  \; \left( S^{-1}_\sigma \right)_{ii'}^{LL'} \;
  \Phi_{i'L'\sigma}^\ast({\bf r}'-{\bf R}_{i'})
  = \delta({\bf r}-{\bf r}') \: .
\label{eq:basis1}
\end{equation}
The wave functions $\Phi_{iL\sigma}({\bf r}-{\bf R}_i)$ are assumed 
to be well localized at the respective sites ${\bf R}_i$ of the 
crystal lattice. $L=(l, m)$ is a composite index characterizing the 
electron's angular momentum, and $S$ denotes the overlap matrix: 

\begin{equation}
  S_{ii'\sigma}^{LL'} =
  \int \Phi_{iL\sigma}^\ast({\bf r}-{\bf R}_{i})
  \Phi_{i'L'\sigma}({\bf r}-{\bf R}_{i'}) d {\bf r} \: .
\label{eq:basis2}
\end{equation}
Since the basis functions are chosen to have definite
angular-momentum character, the basis is non-orthogonal. 
Furthermore, the choice of the basis reflects the relevant energy
range around the Fermi energy we are interested in. To justify the 
usual restriction to the subspace ${\cal H}_S$, the states 
$|iL\sigma\rangle$ are assumed to be orthogonal to all core states
at all sites.

Using the tight-binding representation for the LDA Hamiltonian,

\begin{equation}
  T_{ii'\sigma}^{LL'} = 
  \langle i L \sigma 
  | h_{\rm LDA}|
  i' L' \sigma \rangle
  \: 
\label{eq:hoptb}
\end{equation}
and for the self-energy,

\begin{equation}
  \Sigma_{ii'\sigma}^{LL'}(E_1) = 
  \langle i L \sigma 
   | \Sigma_\sigma(E_1) |
  i' L' \sigma \rangle
  \: ,
\label{eq:sigtb}
\end{equation}
the Dyson equation for the initial-state Green function reads:
 
\begin{equation}
  \sum_{i''L''} \left\{
  E_1 S_{ii''\sigma}^{LL''} 
  - T_{ii''\sigma}^{LL''} 
  - \Sigma_{ii''\sigma}^{LL''}(E_1) \right\}
  G^{(+)L''L'}_{i''i'\sigma}(E_1) 
  = \hbar \delta_{ii'} \delta_{LL'} \: ,
\label{eq:eom}
\end{equation}
Within the tight-binding formulation it can be solved numerically 
by Fourier transformation to 
${\bf k}$ or ${\bf k}_\|$ space and subsequent matrix inversion. 
More important, direct contact is made with numerous many-body 
approaches that likewise refer to a one-particle basis of 
localized orbitals and yield the self-energy in the form of 
Eq.\ (\ref{eq:sigtb}).

For the evaluation of Pendry's formula we consider the system to be
built up from layers parallel to the surface. Because of the damping
of the final-state wave field due to the imaginary part of the inner
potential $V_{0{\rm i}}(E_2)$, it is sufficient
to restrict oneself to a finite number of layers, i.\ e.\ to a slab
geometry. Perfect translational symmetry is assumed within every
layer. We label the layers by an index $i_\perp$. For each 
we define ${\bf R}_{i_\perp}$ to be the position of a reference atom 
which can be thought of as a local origin for the layer $i_\perp$.
Atoms within the layer $i_\perp$ are labeled by an index $i_\|$.
Their positions are given by
${\bf R}_i = {\bf R}_{i_\|} + {\bf R}_{i_\perp}$, where
${\bf R}_{i_\|}$ denotes a vector of the two-dimensional lattice.
(For simplicity we consider a system with one atom per layer unit
cell.)

Two-dimensional translational symmetry is made explicit in the 
notation of the basis orbitals:

\begin{equation}
  | i L \sigma \rangle \equiv
  | i_\| i_\perp L \sigma \rangle \: ,
\label{eq:paraperp}
\end{equation}
from which we construct two-dimensional 
Bloch sums in the following way:

\begin{equation}
  | {\bf q}_\| i_\perp L \sigma \rangle =
  \frac{1}{\sqrt{N_\|}} \sum_{i_\|} 
  e^{i {\bf q}_\| {\bf R}_{i_\|} }
  | i_\| i_\perp L \sigma \rangle
  \: .
\label{eq:blochsum}
\end{equation}
Here ${\bf q}_\|$ is a vector of the first Brillouin zone, and
$N_\|$ denotes the number of atoms within a layer 
($N_\|\mapsto \infty$). 
By means of Fourier transformation to the Bloch-sum basis, the 
Green function can be rewritten as:

\begin{equation}
  G^{(+)LL'}_{{\bf q}_\| i_\perp {i_\perp}'\sigma}(E_1) =
  \frac{1}{N_\|} \sum_{i_\| {i_\|}'}
  e^{-i {\bf q}_\| ({\bf R}_{i_\|}-{\bf R}_{{i_\|}'})}
  G^{(+)LL'}_{ii'\sigma}(E_1) \: .
\label{eq:grkprep}
\end{equation}
The operator representation of the Green function now reads:

\begin{equation}
  G_{1\sigma}^{+} = 
  \sum_{{\bf q}_\| i_\perp {i_\perp}' LL'}
  | {\bf q}_\| i_\perp L \sigma \rangle 
  \; G^{(+)LL'}_{{\bf q}_\| i_\perp {i_\perp}'\sigma}(E_1) \;
  \langle {\bf q}_\| {i_\perp}' L' \sigma | 
  \: .
\label{eq:oprep}
\end{equation}
The operator $G_{1\sigma}^{+}$ is independent from the choice of 
the one-particle basis. Inserting into Pendry's formula, we get:

\begin{equation}
  I^{\rm PES} \propto f_{\rm F}(E_1) \:
  \mbox{Im} \sum_{i_\perp {i_\perp}'LL'} \sum_{{\bf q}_\|}
  M_{i_\perp L\sigma} (\epsilon_f, {\bf k}_\|, {\bf q}_\|) \:
  G^{(+)LL'}_{{\bf q}_\| i_\perp {i_\perp}'\sigma}(E_1) \;
  M_{{i_\perp}' L'\sigma}^\ast (\epsilon_f, {\bf k}_\|, {\bf q}_\|) 
  \: ,
\label{eq:iphototme}
\end{equation}
where $M$ is the matrix element of the dipol operator 
between the final state 
$|f\rangle = G_{2\sigma}^- |\epsilon_f {\bf k}_\| \sigma \rangle$
and the Bloch sum $|{\bf q}_\| i_\perp L\sigma\rangle$:

\begin{equation}
  M_{i_\perp L\sigma} (\epsilon_f, {\bf k}_\|, {\bf q}_\|) =
  \langle \epsilon_f {\bf k}_\| \sigma | G_{2\sigma}^+ 
  \Delta |{\bf q}_\| i_\perp L\sigma \rangle \: .
\label{eq:tmedef}
\end{equation}

The final state $G_{2\sigma}^- |\epsilon_f {\bf k}_\| \sigma \rangle$
is an eigenstate of $h_{\rm LDA}$ with eigenenergy $E_2=\epsilon_f$.
Let us also consider the low-energy 
eigenstates of $h_{\rm LDA}$ within
the subspace ${\cal H}_S$ which we refer to in the following as the 
initial states. These may be characterized by the spin
projection $\sigma$ and the parallel wave vector ${\bf q}_\|$ which
are good quantum numbers: 

\begin{equation}
  h_{\rm LDA} | n {\bf q}_\| \sigma \rangle =
  \epsilon_{n\sigma}({\bf q}_\|) | n {\bf q}_\| \sigma \rangle 
  \: .
\label{eq:ldaeigvalproblem}
\end{equation}
Here $\epsilon_{n\sigma}({\bf q}_\|)$ are the LDA eigenenergies. The 
LDA eigenstates form an orthonormal basis of ${\cal H}_S$. We make 
use of them to rewrite the transition-matrix element:

\begin{equation}
  M_{i_\perp L\sigma}(\epsilon_f, {\bf k}_\|, {\bf q}_\|) = 
  \sum_{n}
  \langle \epsilon_f {\bf k}_\| \sigma | G_{2\sigma}^+ \Delta
  | n {\bf q}_\| \sigma \rangle
  \langle  n {\bf q}_\| \sigma |
  {\bf q}_\| i_\perp L\sigma \rangle \: .
\label{eq:tmewithone}
\end{equation}
Thereby the matrix element of the dipole operator is given between
eigenstates of $h_{\rm LDA}$. 
This is shown below (Sec.\ \ref{sec:tme})
to facilitate the calculation to a great extent.

We now proceed as follows: In the next section we will deal with the
initial states $|n{\bf q}_\|\sigma \rangle$, Sec.\ \ref{sec:fin} 
focusses on the final state $|f\rangle$ until finally we are in the
position to derive a computationally feasible expression for the
transition-matrix elements in Sec.\ \ref{sec:tme}.

\section{THE INITIAL STATES}
\label{sec:ini}

Let us ``switch off'' the non-local self-energy for a moment, 
i.~e.\ take $\Sigma \equiv 0$. In this case the Green function 
becomes diagonal in the orthonormal basis of the initial states:

\begin{equation}
  G_{1\sigma}^{+} = 
  \sum_{n {\bf q}_\|} | n {\bf q}_\| \sigma \rangle
  G^{+}_{n {\bf q}_\| \sigma}(E_1) \langle n {\bf q}_\| \sigma |
  \: .
\end{equation}
We then have

\begin{equation}
  - \frac{1}{\pi} \mbox{Im} G^{+}_{n {\bf q}_\| \sigma}(E_1) =
  \hbar \delta( E_1-(\epsilon_{n\sigma}({\bf q}_\|) - \mu) )
\end{equation}
as a factor within Pendry's formula (\ref{eq:pendry}). For the case
$\Sigma \equiv 0$ we can thus conclude that it is 
sufficient to consider merely a single initial state, namely the 
one with the eigenenergy 
$\epsilon_{n\sigma}({\bf k}_\|) = E_1+\mu = 
\epsilon_{f\sigma}({\bf k}_\|) - \hbar \omega$. If the self-energy 
is switched on again, its imaginary part will cause a broadening of 
the delta-function indicating a finite quasi-particle lifetime. In a
sense the LDA band structure will be smeared out to some degree.
Furthermore, strong correlations may lead to satellites in the 
excitation spectrum which take some spectral weight from the 
main bands. Both effects imply that for a given photon
energy $\hbar \omega$ there may be different initial states 
$| n {\bf q}_\| \sigma \rangle$ coupled to the same final 
state $|f\rangle$. Opposed to the conventional evaluation 
of the one-step model, we thus need all eigenstates of 
$h_{\rm LDA}$ around the energy 
$\epsilon_{f\sigma}({\bf k}_\|)-\hbar \omega$.

We expand the initial states in the Bloch-sum basis:

\begin{equation}
  | n {\bf q}_\| \sigma \rangle =
  \sum_{i_\perp L} \alpha_{i_\perp L \sigma}^{n {\bf q}_\|}
  | {\bf q}_\| i_\perp L \sigma \rangle
  \: .
\label{eq:ldaeigstates}
\end{equation}
The expansion coefficients 
$\alpha_{{i_\perp} L \sigma}^{n{\bf q}_\|}$ as well 
as the LDA eigenenergies have to be determined from the 
following non-orthogonal eigenvalue problem:

\begin{equation}
  \sum_{{i_\perp}'L'} \left( 
  T_{i_\perp {i_\perp}' \sigma}^{LL'} ({\bf q}_\|) -
  \epsilon_{n\sigma} ({\bf q}_\|) 
  S_{i_\perp {i_\perp}' \sigma}^{LL'} ({\bf q}_\|)
  \right)
  \alpha_{{i_\perp}' L' \sigma}^{n{\bf q}_\|} = 0
  \: ,
\label{eq:ldacoeff}
\end{equation}
which is easily derived from (\ref{eq:ldaeigvalproblem}) when 
expressing the LDA-Hamiltonian and the overlap within the Bloch-sum
basis:

\begin{eqnarray}
  T_{i_\perp {i_\perp}' \sigma}^{LL'} ({\bf q}_\|) & = &
  \langle {\bf q}_\| i_\perp L \sigma | h_{\rm LDA} |
  {\bf q}_\| {i_\perp}' L' \sigma \rangle \: ,
  \nonumber \\
  S_{i_\perp {i_\perp}' \sigma}^{LL'} ({\bf q}_\|) & = &
  \langle {\bf q}_\| i_\perp L \sigma | 
  {\bf q}_\| {i_\perp}' L' \sigma \rangle
  \: .
\label{eq:tqsq}
\end{eqnarray}
The ${\bf q}_\|$-dependent (LDA-)Hamilton and overlap matrices are 
connected with the Hamilton and overlap matrices from Eqs.
(\ref{eq:basis2}) and (\ref{eq:hoptb}) via two-dimensional 
Fourier transformation:

\begin{eqnarray}
  T_{ii'\sigma}^{LL'} & = &
  \frac{1}{N_\|} \sum_{{\bf q}_\|}
  e^{i {\bf q}_\| ({\bf R}_{i_\|}-{\bf R}_{i_\|}')}
  T_{i_\perp {i_\perp}' \sigma}^{LL'} ({\bf q}_\|) \: ,
  \nonumber \\
  S_{ii'\sigma}^{LL'} & = &
  \frac{1}{N_\|} \sum_{{\bf q}_\|}
  e^{i {\bf q}_\| ({\bf R}_{i_\|}-{\bf R}_{i_\|}')}
  S_{i_\perp {i_\perp}' \sigma}^{LL'} ({\bf q}_\|)
  \: .
\label{eq:tsqtsi}
\end{eqnarray}
From Eqs.\ (\ref{eq:ldaeigstates}) and (\ref{eq:tqsq}) we have:

\begin{equation}
  \langle n {\bf q}_\| \sigma |{\bf q}_\| i_\perp L\sigma \rangle 
  =
  \sum_{{i_\perp}'L'}
  \left( \alpha_{{i_\perp}'L'\sigma}^{n{\bf q}_\|} \right)^\ast
  S_{{i_\perp}' i_\perp \sigma}^{L'L}({\bf q}_\|)
  \: ,
\label{eq:nqil}
\end{equation}
which gives us the second factor in Eq.\ (\ref{eq:tmewithone}) for 
the transition-matrix elements.

For the first factor, i.\ e.\ for the actual matrix element, we need
the real-space representation of the initial state 
$|n {\bf q}_\| \sigma \rangle$:

\begin{equation}
  \Psi^{\rm (n)}_{{\bf q}_\|\sigma}({\bf r}) \equiv 
  \langle {\bf r} | n {\bf q}_\| \sigma \rangle = 
  \sum_{i_\perp L} \alpha_{{i_\perp}L\sigma}^{n{\bf q}_\|}
  \Phi^{{\bf q}_\|}_{i_\perp L\sigma}({\bf r}-{\bf R}_{i_\perp}) \: .
\label{eq:psirealspace}
\end{equation}
Here we have introduced the real-space representation of the Bloch
sums [cf.\ Eqs.\ (\ref{eq:basis}) and (\ref{eq:blochsum})]:

\begin{equation}
  \Phi^{{\bf q}_\|}_{i_\perp L\sigma}({\bf r}-{\bf R}_{i_\perp})
  \equiv 
  \langle {\bf r} | {\bf q}_\| i_\perp L \sigma \rangle
  = 
  \frac{1}{\sqrt{N_\|}} \sum_{i_\|} 
  e^{i {\bf q}_\| {\bf R}_{i_\|}}
  \Phi_{iL\sigma} ({\bf r}-{\bf R}_i) \: .
\label{eq:blochrealspace}
\end{equation}

To proceed further, it becomes necessary to specify the basis 
orbitals $\Phi_{iL\sigma}({\bf r})$ and the form of the LDA 
potential. We assume that $V_{\rm LDA}({\bf r})$ can be 
sufficiently well approximated by a muffin-tin potential, 
which is spherically symmetric within non-overlapping muffin-tin 
spheres centered at the lattice sites ${\bf R}_i$ and constant 
inbetween. The muffin-tin approximation will greatly facilitate 
the calculation of the transition-matrix elements since we then 
have a simple separation in radial and angular parts. This also 
implies the need for a one-center expansion of the initial states 
(\ref{eq:psirealspace}) and thus of the Bloch sums, which in 
Eq.\ (\ref{eq:blochrealspace}) are given in terms of the basis 
orbitals centered at all sites $i_\|$ within a layer $i_\perp$. 
For this purpose we may choose the basis orbitals to be 
(augmented) muffin-tin orbitals (MTO's) \cite{Skr84,SS91}.
Being orthogonal to all core states at all sites of the lattice
and spanning the subspace ${\cal H}_S$, they have all the properties
we demanded above. Furthermore, we have the desired one-center
expansion for the Bloch sum of the MTO's of a layer $i_\perp$:

\begin{equation}
  \sqrt{N_\|} \; 
  \Phi^{{\bf q}_\|}_{i_\perp L\sigma} ({\bf r}-{\bf R}_{i_\perp})
  = \delta_{{i_\perp}' i_\perp} 
  \Phi_{i L \sigma} ({\bf r}-{\bf R}_{{i_\perp}'}) -
  \sum_{L'} Z^{{i_\perp}' i_\perp}_{L'L}({\bf q}_\|) \;
  \widetilde{\Phi}_{i L' \sigma} ({\bf r}-{\bf R}_{{i_\perp}'}) \: .
\label{eq:onecenter}
\end{equation}
The expansion converges within the muffin-tin sphere at 
${\bf R}_{{i_\|}'}=0$ of a given layer ${i_\perp}'$ \cite{Skr84}.
$Z$ is the potential-independent structure constant and
$\widetilde{\Phi}$ an augmented spherical Bessel function. Let us 
mention that the existence of the one-center expansion rests on an 
analogous expansion theorem for the spherical Neumann functions
and not necessarily on the muffin-tin form of the LDA potential
\cite{SS91}. $\Phi$ and $\widetilde{\Phi}$ are separated into a
radial and an angular part:

\begin{eqnarray}
  \Phi_{i L\sigma} ({\bf r}) & = & 
  \phi_{il\sigma} (r) Y_L(\hat{\bf r}) \: ,
  \nonumber \\
  \widetilde{\Phi}_{iL\sigma} ({\bf r}) & = & 
  \widetilde{\phi}_{il\sigma} (r) Y_L(\hat{\bf r}) \: .
\label{eq:radang}
\end{eqnarray}
Because of two-dimensional translational symmetry within a layer, 
the radial parts are the same for all sites $i_\|$ within a layer 
and depend on the layer index $i_\perp$ only. A further
specification of the radial parts is not necessary for our present
purposes.

Inserting (\ref{eq:onecenter}) and (\ref{eq:radang}) into 
(\ref{eq:psirealspace}) we get

\begin{equation}
  \Psi^{(n)}_{{\bf q}_\|\sigma}({\bf r}) =
  \frac{1}{\sqrt{N_\|}} \sum_{L}
  \left[
  B_{i_\perp L \sigma}^{n{\bf q}_\|} \;
  \phi_{i_\perp l \sigma} (|{\bf r}-{\bf R}_{i_\perp}|)
  + 
  \widetilde{B}_{i_\perp L \sigma}^{n{\bf q}_\|} \;
  \widetilde{\phi}_{i_\perp l \sigma} (|{\bf r}-{\bf R}_{i_\perp}|)
  \right]
  Y_{L}(\widehat{{\bf r}-{\bf R}_{i_\perp}}) \: ,
\label{eq:infinal}
\end{equation}
where we have defined

\begin{eqnarray}
  B_{i_\perp L \sigma}^{n{\bf q}_\|} & = &
  \alpha_{{i_\perp} L \sigma}^{n{\bf q}_\|} \: ,
  \nonumber \\
  \widetilde{B}_{i_\perp L \sigma}^{n{\bf q}_\|} & = &
  \sum_{{i_\perp}' L'} Z^{i_\perp {i_\perp}'}_{LL'}({\bf q}_\|)
  \; \alpha_{{i_\perp}' L' \sigma}^{n{\bf q}_\|}  \: .
\label{eq:bdef}
\end{eqnarray}
Therewith we have the final one-center expansion of the initial 
states into spherical harmonics.

\section{THE FINAL STATE}
\label{sec:fin}

For the determination of the final state

\begin{equation}
  \Psi^{\rm (f)}_{{\bf k}_\| \sigma}({\bf r}) = 
  \langle {\bf r} | G_{2\sigma}^- | 
  \epsilon_f {\bf k}_\| \sigma \rangle 
\label{eq:finspace}
\end{equation}
we consider a conventional diffraction situation with 
$|\epsilon_f {\bf k}_\| \sigma \rangle$ (the state in the
vacuum at the photoelectron detector) interpreted as a {\em source}
of electrons which penetrate into the crystal. Within the framework
of the layer-KKR multiple-scattering formalism 
\cite{Kor47,KR54,Kam67,Kam68,Pen74} we can construct
the whole high-energy wave field from the source
$|\epsilon_f {\bf k}_\| \sigma \rangle$ and from the
scattering properties of each layer. The result 
gives us the {\em time-reversed} final state 
$\left( \Psi^{\rm (f)}_{{\bf k}_\| \sigma}({\bf r}) \right)^\ast$.
Just as in the case of the initial states the main goal is to derive
a one-center expansion of the final state into spherical harmonics
within a given muffin-tin sphere at 
${\bf r}={\bf R}_{i_\|}+{\bf R}_{i_\perp}$. 

To begin with, we consider the region of constant inner potential 
$V_{0,2}$ between the muffin-tin spheres where plane waves are 
solutions of Schr\"odinger's equation. The potential $V_{\rm LDA}$ 
can be assumed to be invariant under two-dimensional lattice 
translations ${\bf r} \mapsto {\bf r} + {\bf R}_{i_\|}$. For any 
multiple-scattering event this implies that the wave vector parallel 
to the surface ${\bf k}_\|$ can be changed by a reciprocal lattice 
vector ${\bf g}_\|$ only. In the interstitial region the high-energy 
wave field can thus be expanded into plane waves with wave vectors

\begin{equation}
  {\bf k}_{{\bf g}_\|}^{\pm} = 
  \left(
  - {\bf k}_\| + {\bf g}_\| \: , \:
  \pm \sqrt{
  2 m_e (E_2 - V_{0,2}) / \hbar^2 - |-{\bf k}_\| + {\bf g}_\| |^2
  } \:
  \right) \: .
\end{equation}

We define the positive $z$-axis pointing into the crystal
and choose ${\bf r}={\bf R}_{i_\perp}$ to be a local origin for a 
given layer $i_\perp$. Consider then a plane-wave field advancing 
on the layer from the $+z$ side:

\begin{equation}
  \frac{1}{\sqrt{N_\|}}
  \sum_{{\bf g}_\|} 
  W^+_{{\bf g}_\| i_\perp \sigma}
  e^{i {\bf k}_{{\bf g}_\|}^+ ({\bf r}-{\bf R}_{i_\perp})} \: .
\end{equation}
Scattering at the layer $i_\perp$ gives rise to a transmitted and to
a reflected wave field on the $+z$ and the $-z$ side of the layer 
with coefficients $V^+_{{\bf g}_\| i_\perp \sigma}$ and
$W^-_{{\bf g}_\| i_\perp \sigma}$, respectively.
They are given via the transmission and reflection matrices:

\begin{eqnarray}
  V^+_{{\bf g}_\| i_\perp \sigma} & = & 
  \sum_{{\bf g}'_\|}
  T_{{\bf g}_\| {\bf g}'_\|}^{i_\perp \sigma} 
  W^+_{{\bf g}'_\| i_\perp \sigma} \: ,
  \nonumber \\
  W^-_{{\bf g}_\| i_\perp \sigma} & = & 
  \sum_{{\bf g}'_\|}
  R_{{\bf g}_\| {\bf g}'_\|}^{i_\perp \sigma} 
  W^+_{{\bf g}'_\| i_\perp \sigma} \: .
\end{eqnarray}
Similar relations hold for an advancing wave field from the $-z$ 
side. Conventional LEED theory \cite{Pen74} provides us with 
explicit expressions for the scattering matrices, let us
cite the final result:

\begin{eqnarray}
  T_{{\bf g}_\| {\bf g}'_\|}^{i_\perp \sigma} 
  & = & 
  \frac{8 \pi^2}{\kappa A |k^+_{{\bf g}_\| z}|} 
  \sum_{LL'} \: i^{-l} \: 
  Y_L(\widehat{{\bf k}^+_{{\bf g}'_\|}}) \:
  t_{l i_\perp \sigma} \:
  ({\bf 1} - {\bf X})^{-1}_{LL'} \:
  i^{l'} \: (-1)^{m'} 
  Y_{\overline{L'}}(\widehat{{\bf k}^+_{{\bf g}'_\|}})
  + \delta_{{\bf g}_\| {\bf g}'_\|} \: ,
  \nonumber \\
  R_{{\bf g}_\| {\bf g}'_\|}^{i_\perp \sigma} 
  & = & 
  \frac{8 \pi^2}{\kappa A |k^+_{{\bf g}_\| z}|} 
  \sum_{LL'} \: i^{-l} \: 
  Y_L(\widehat{{\bf k}^+_{{\bf g}'_\|}}) \:
  t_{l i_\perp \sigma} \:
  ({\bf 1} - {\bf X})^{-1}_{LL'} \:
  i^{l'} \: (-1)^{m'} 
  Y_{\overline{L'}}(\widehat{{\bf k}^-_{{\bf g}'_\|}}) 
  \: .
\label{eq:tandr}
\end{eqnarray}
Here we have written 
$t_{li_\perp \sigma}=(e^{2i\delta_{li_\perp \sigma}}-1)/2$
for abbreviation. By $\delta_{li_\perp \sigma}$ we denote the 
layer- and spin-dependent phase shifts which completely characterize 
the scattering properties of the spherically symmetric potential 
within a single muffin-tin sphere. Furthermore, $A$ stands for the
area of the layer unit cell, 
$\kappa=\sqrt{2m_e (E_2-V_{0,2})}/\hbar$ and $\overline{L}=(l,-m)$.
The matrix 

\begin{equation}
  X_{LL'} = t_{l i_\perp \sigma} {\sum_{i_\|L''} }'
  4 \pi i^{(l-l'-l'')} 
  h_{l''}^{(1)}(-{\bf k}_\|{\bf R}_{i_\|}) \:
  Y_{\overline{L''}}(\widehat{-{\bf R}_{i_\|}})
  (-1)^{(m'+m'')} C_{L\overline{L'}L''} \:
  e^{i{\bf k}_\|{\bf R}_{i_\|}} \: ,
\label{eq:xmat}
\end{equation}
which is due to Kambe \cite{Kam67,Kam68}, corrects for intra-layer
multiple-scattering events. Here $h_l^{(1)}$ denotes the spherical 
Hankel function and 

\begin{equation}
  C_{LL'L''} = \int_{(4\pi)} 
  Y_{L}(\Omega) \: Y_{L'}(\Omega) \: Y_{L''}(\Omega) \: d\Omega 
\label{eq:gaunt}
\end{equation}
are the Gaunt coefficients.
The definition of the $X$-matrix is closely related to the structure 
constants defined via Eq.\ (\ref{eq:onecenter}) \cite{Skr84}.

By the matrices (\ref{eq:tandr}) the scattering properties of all  
layers are known. For a given plane wave 
$|\epsilon_f {\bf k}_\| \sigma \rangle$ advancing onto the crystal
from the vacuum side, they determine the whole high-energy wave field
between all layers in the semi-infinite crystal. Its coefficients
can straightforwardly be found within a recursive layer-by-layer
scheme \cite{Pen76}. Since the 
flux of elastically scattered electrons
is permanently reduced due to the imaginary part of the (generalized)
inner potential, the wave field 
can be neglected beyond a finite depth.

Once all coefficients of the plane-wave expansions are known, the 
next step will be to derive an expansion of the high-energy wave
field into spherical harmonics centered at the local origin 
${\bf R}_{i_\perp}$ of a layer $i_\perp$. The advancing plane-wave
field at the layer $i_\perp$ is:

\begin{equation}
  \Psi^{(0)}_{{\bf k}_\| i_\perp \sigma} ({\bf r}) =
  \frac{1}{\sqrt{N_\|}}
  \sum_{{\bf g}_\|} 
  \left(
  W^+_{{\bf g}_\| i_\perp \sigma}
  e^{i {\bf k}_{{\bf g}_\|}^+ ({\bf r}-{\bf R}_{i_\perp})}
  +
  V^-_{{\bf g}_\| i_\perp \sigma}
  e^{i {\bf k}_{{\bf g}_\|}^- ({\bf r}-{\bf R}_{i_\perp})}
  \right) \: ,
\label{eq:advwf}
\end{equation}
where $W^+_{{\bf g}_\| i_\perp \sigma}$ and 
$V^-_{{\bf g}_\| i_\perp \sigma}$ are the coefficients
of the plane waves on the $-z$ and the $+z$ side of the 
layer, respectively. From (\ref{eq:advwf}) we get the spherical
wave amplitudes

\begin{equation}
  A_{i_\perp L \sigma}^{{\bf k}_\| (0)} =
  \frac{1}{\sqrt{N_\|}}
  \sum_{{\bf g}_\|} 4\pi i^l \: (-1)^m 
  \left(  
  W^+_{{\bf g}_\| i_\perp \sigma} 
  Y_{\overline{L}}(\widehat{{\bf k}_{{\bf g}_\|}^+})
  +
  V^-_{{\bf g}_\| i_\perp \sigma} 
  Y_{\overline{L}}(\widehat{{\bf k}_{{\bf g}_\|}^-})
  \right)
\label{eq:advwf1}
\end{equation}
of the advancing wave field:

\begin{equation}
  \Psi^{(0)}_{{\bf k}_\| i_\perp \sigma} ({\bf R}_{i_\perp}+{\bf r}) 
  = \sum_L
  A_{i_\perp L \sigma}^{{\bf k}_\| (0)} \;
  j_l(\kappa r) Y_L(\Omega) \: ,
\end{equation}
where $j_l$ is the spherical Bessel function. Correcting for 
multiple scattering within the layer \cite{Pen76},

\begin{equation}
  A_{i_\perp L \sigma}^{{\bf k}_\|} = \sum_{L' }
  A_{i_\perp L' \sigma}^{{\bf k}_\| (0)} \:
  ({\bf 1} - {\bf X})^{-1}_{L'L} \: ,
\label{eq:advwf2}
\end{equation}
we arrive at the total final-state wave field inside the muffin-tin
sphere at ${\bf R}_{i_\perp}$:

\begin{equation}
  \left( \Psi^{(f)}_{{\bf k}_\| \sigma} ({\bf r}) \right)^\ast =
  \frac{1}{\sqrt{N_\|}} \sum_L
  A_{i_\perp L \sigma}^{{\bf k}_\|} \:
  e^{i \delta_{l i_\perp \sigma}} \:
  \psi_{i_\perp l \sigma}(|{\bf r}- {\bf R}_{i_\perp}|) \:
  Y_L(\widehat{{\bf r}-{\bf R}_{i_\perp}}) \: .
\label{eq:finfinal}
\end{equation}
$\psi_{i_\perp l \sigma}(r)$ is the radial part of the solution 
of the Schr\"odinger equation for energy $E_2$ that is regular 
at the origin. The radial function as well as the phase shifts 
have to be determined numerically for the spherically symmetric 
LDA potential within the muffin-tin sphere at ${\bf R}_{i_\perp}$.

\section{THE TRANSITION-MATRIX ELEMENTS}
\label{sec:tme}

We are now in a position that allows for calculating the actual 
transition-matrix elements, i.\ e.\ the first factor in 
Eq.\ (\ref{eq:tmewithone}). Again we turn to real-space
representation:

\begin{equation}
  \langle \epsilon_f {\bf k}_\| \sigma | G_{2\sigma}^+ \Delta
  | n {\bf q}_\| \sigma \rangle =
  \frac{-i\hbar e}{m_e} 
  \int \left( \Psi^{\rm (f)}_{{\bf k}_\| \sigma} ({\bf r}) 
  \right)^\ast
  \left( {\bf A}_0 \nabla \right) 
  \Psi^{\rm (n)}_{{\bf q}_\|\sigma}({\bf r})
  d{\bf r} \: .
\label{eq:menabla}
\end{equation}
Since 
$[\nabla ,h_{\rm LDA}]_- =\nabla V_{{\rm LDA},\sigma}({\bf r})$,
the transition-matrix element can be rewritten as:

\begin{equation}
  \langle \epsilon_f {\bf k}_\| \sigma | G_{2\sigma}^+ \Delta
  | n {\bf q}_\| \sigma \rangle =
  \frac{i\hbar e}{m_e} \:
  \frac{1}{\epsilon_{f\sigma}({\bf k}_\|) - 
  \epsilon_{n\sigma}({\bf q}_\|)}
  \int \! \left( \Psi^{\rm (f)}_{{\bf k}_\| \sigma} ({\bf r})
  \right)^\ast
  \left( {\bf A}_0 \nabla V_{{\rm LDA},\sigma}({\bf r})\right) 
  \Psi^{\rm (n)}_{{\bf q}_\|\sigma}({\bf r})
  d{\bf r} \: .
\label{eq:mecomm}
\end{equation}
Here we made use of the fact that both, the final state 
$\Psi^{\rm (f)}$ and the initial states $\Psi^{\rm (n)}$,
are eigenstates of $h_{\rm LDA}$. At this point we thus profit
from the transformation of the 
transition-matrix element (\ref{eq:tmedef})
into the expression given by Eq.\ (\ref{eq:tmewithone}). Since
$\nabla V_{{\rm LDA},\sigma}({\bf r}) \equiv 0$ in the region of 
constant inner potential between the muffin-tin spheres, the integral
in (\ref{eq:mecomm}) now reduces to a sum of integrals over all 
muffin-tin spheres:

\begin{eqnarray}
  \nabla V_{{\rm LDA},\sigma} ({\bf r}) & = &
  \sum_i \Theta_{\rm MT}({\bf r}-{\bf R}_i)
  \nabla V_{{\rm LDA},\sigma} ({\bf r}) \nonumber \\
  & \equiv & \sum_i \nabla V_{i\sigma}({\bf r}-{\bf R}_i) \: ,
\label{eq:vdecomp}
\end{eqnarray}
where $\Theta_{\rm MT}({\bf r})=1$ for $r<r_{\rm MT}$ and 
$\Theta_{\rm MT}({\bf r})=0$ for $r>r_{\rm MT}$, and $r_{\rm MT}$ 
is the radius of the muffin-tin sphere.
$V_{i\sigma}({\bf r})=V_{i\sigma}(r)$ is the potential within the
sphere at the site $i$ with $V_{i\sigma}(r)=0$ for $r>r_{\rm MT}$.

Another considerable simplification of the transition-matrix elements
arises from the two-dimensional lattice periodicity of the LDA
potential, $V_{i\sigma}=V_{i_\perp\sigma}$: Since both, the final
state and the initial states, fulfill the two-dimensional analogue
of Bloch's theorem, any lattice translation of the form
${\bf r} \mapsto {\bf r} + {\bf R}_{i_\|}$ leaves the integral
(\ref{eq:mecomm}) invariant apart from a factor 
${\rm exp}(i({\bf q}_\|-{\bf k}_\|){\bf R_{i_\|}})$.
We thus can conclude that the difference ${\bf q}_\|-{\bf k}_\|$
is equal to a reciprocal lattice vector ${\bf g}_\|$. Since we can
consider ${\bf k}_\|$ to be fixed by the energy of the photoelectron
at the detector and by the emission angles, the reciprocal lattice
vector is uniquely determined by demanding ${\bf q}_\|$ to lie 
within the first surface Brillouin zone. 
Hence, translational symmetry reduces
the ${\bf q}_\|$-sum in Eq.\ (\ref{eq:iphototme}) to a single term:

\begin{equation}
  {\bf q}_\| = {\bf k}_\| + {\bf g}_\| \: .
\label{eq:qparallel}
\end{equation}
Referring once more to translational symmetry, we also conclude that
all integrals over muffin-tin spheres within the same layer $i_\perp$
are equal. This leads us finally to:

\begin{eqnarray}
  \langle \epsilon_f {\bf k}_\| \sigma | G_{2\sigma}^+ \Delta
  | n ({\bf k}_\| + {\bf g}_\|) \sigma \rangle & = &
  \\ && \hspace{-45mm} 
  N_\| \; \frac{i\hbar e}{m_e} \:
  \frac{1}{\epsilon_{f\sigma}({\bf k}_\|) - 
  \epsilon_{n\sigma}({\bf k}_\|+{\bf g}_\|)}
  \int_{S(i_\perp)}
  \left( \Psi^{\rm (f)}_{{\bf k}_\| \sigma} ({\bf r}) \right)^\ast
  {\bf A}_0 \nabla V_{i_\perp \sigma}({\bf r}-{\bf R}_{i_\perp}) 
  \Psi^{\rm (n)}_{{\bf k}_\| + {\bf g}_\| \sigma}({\bf r})
  d{\bf r} \: , \nonumber 
\end{eqnarray}
where the integral extends over the sphere at the local origin 
${\bf R}_{i_\perp}$ of the layer $i_\perp$.

For the evaluation of the integral we separate into radial and 
angular parts. For this purpose we need the expansion
of the dipole operator into spherical harmonics:

\begin{eqnarray}
  {\bf A}_0 \nabla V_{i_\perp \sigma}(r) & = &
  \frac{d V_{i_\perp \sigma}}{d r}(r) \: A_0 \:
  \frac{{\bf A}_0{\bf r}}{A_0 r}
  \nonumber \\ & = &
  \frac{d V_{i_\perp \sigma}}{d r}(r) \: A_0 \:
  \frac{4\pi}{3} \sum_{m=-1}^1 
  Y_{1m}^\ast (\widehat{{\bf A}_0}) Y_{1m}(\hat{\bf r}) 
\end{eqnarray}
and the one-center expansions of the initial states 
(\ref{eq:infinal}) and the final state (\ref{eq:finfinal})
as derived in Secs.\ \ref{sec:ini} and \ref{sec:fin}. The 
integration over the angular parts can be performed analytically 
and yields the following angular matrix elements:

\begin{equation}
  D_{LL'} = \frac{4\pi}{3} A_0 \: Y_{1m}^\ast(\widehat{{\bf A}_0}) \:
  C_{lml'm'1(-m-m')} \: ,
\end{equation}
where we have used the definition (\ref{eq:gaunt}) of the Gaunt 
coefficients. The integration over the radial parts results
in two types of radial matrix elements:
 
\begin{eqnarray}
  M_{i_\perp ll'\sigma} = e^{i \delta_{l i_\perp \sigma}} 
  \int_0^{r_{\rm MT}} r^2
  \psi_{i_\perp l \sigma}(r)
  \frac{dV_{i_\perp \sigma}}{dr}(r)
  \phi_{i_\perp l' \sigma}(r) dr \: ,
  \nonumber \\
  \widetilde{M}_{i_\perp ll'\sigma} = e^{i \delta_{l i_\perp \sigma}}
  \int_0^{r_{\rm MT}} r^2
  \psi_{i_\perp l \sigma}(r)
  \frac{dV_{i_\perp \sigma}}{dr}(r)
  \widetilde{\phi}_{i_\perp l' \sigma}(r) dr \: ,
\label{eq:rme}
\end{eqnarray}
in the definition of which we included the phase shifts from 
Eq.\ (\ref{eq:finfinal}). Since $\psi_{i_\perp l \sigma}(r)$ as well 
as $\phi_{i_\perp l \sigma}(r)$ and 
$\widetilde{\phi}_{i_\perp l \sigma}(r)$ are regular functions
at $r=0$ and since 
$V_{i_\perp \sigma}(r) \propto -Z/r$ for $r \mapsto 0$,
the integrals are well defined.

The final formula for the photocurrent now reads:

\begin{equation}
  I^{\rm PES} \propto f_{\rm F}(E_1) \:
  \mbox{Im} \sum_{i_\perp {i_\perp}'} \sum_{LL'}
  M_{i_\perp L\sigma} (\epsilon_f, {\bf k}_\|) \:
  G^{(+)LL'}_{{\bf k}_\| + 
  {\bf g}_\| i_\perp {i_\perp}'\sigma}(E_1) \;
  M_{{i_\perp}' L'\sigma}^\ast (\epsilon_f, {\bf k}_\|) \: ,
\label{eq:phofinal}
\end{equation}
where

\begin{eqnarray}
  M_{i_\perp L\sigma} (\epsilon_f, {\bf k}_\|) 
  & = & 
  \frac{ie\hbar}{m_e}
  \sum_n
  \frac{1}{\epsilon_{f\sigma}({\bf k}_\|) - 
  \epsilon_{n\sigma}({\bf k}_\|+{\bf g}_\|)} \times
  \nonumber \\
  && \hspace{-10mm} 
  \left(
  \sum_{{i_\perp}'L'}
  \left( 
  \alpha^{n{\bf k}_\| + {\bf g}_\|}_{{i_\perp}'L'\sigma}
  \right)^\ast
  S_{{i_\perp}' i_\perp \sigma}^{L'L}({\bf k}_\|+{\bf g}_\|)
  \right) \times
  \nonumber \\
  && \hspace{-10mm} 
  \left(
  \sum_{{i_\perp}'} \sum_{LL'} D_{LL'}
  A^{{\bf k}_\|}_{{i_\perp}' L \sigma}
  \left(
  M_{{i_\perp}' ll'\sigma} 
  B^{n{\bf k}_\|+{\bf g}_\|}_{{i_\perp}' L' \sigma}
  +  
  \widetilde{M}_{{i_\perp}' ll'\sigma}
  \widetilde{B}^{n{\bf k}_\|+{\bf g}_\|}_{{i_\perp}' L' \sigma}
  \right)
  \right) \: .
\label{eq:mefinal}
\end{eqnarray}
This completes the formalism. Combining the essentials of the 
previous sections, Eqs.\ (\ref{eq:phofinal}) and (\ref{eq:mefinal}) 
express the result in a rather compact form.

\section{CONCLUDING REMARKS}
\label{sec:con}

This paper has proposed a theory of photoemission (and inverse
photoemission) from single-crystal surfaces that keeps the basic
structure of the well-known one-step model but determines the
initial-state Green function from the Dyson equation rather
than by means of KKR multiple-scattering theory. This approach
requires to reconsider the calculation of the transition-matrix
elements which has been worked out in detail.

In our opinion it is an important advantage of the presented 
reformulation that the main physical concepts inherent in the 
one-step model show up in a very transparent way:

i) The structure of the final equation (\ref{eq:phofinal}) for the 
photocurrent is still reminiscent of first-order perturbation theory 
in the external electric field (Fermi's golden rule). 

ii) Consequently, one can easily distinguish between the ``bare'' 
spectrum which essentially is given by the Green function on the 
one hand and its modifications due to secondary effects arising 
from the wave-vector and energy dependence of the transition-matrix 
elements on the other. 

iii) Due to the Fermi-function cut off (for $T=0$), a non-zero 
intensity is found for energies $E_1<0$ only, i.~e.\ for $\epsilon_f 
- \hbar \omega < \mu$: The initial-state Green function in 
(\ref{eq:phofinal}) describes {\em hole} propagation, namely 
propagation of the residual hole in the valence band that 
corresponds to the excited photoelectron. Since we applied 
the sudden approximation, photoelectron and hole propagate 
independently from each other in time. However, energy 
conservation is satisfied, and thus the photoelectron takes 
away the information on the hole-excitation spectrum. The energy 
dependence of the photoemission spectrum originates from the 
energy dependence of the initial-state (hole) Green function and 
is shifted by the photon energy $\hbar \omega$.

iv) The theory is able to account for electron-correlation effects 
{\em in the initial state}; in (\ref{eq:phofinal}) $G_{1\sigma}^+$ is
the {\em fully interacting} Green function. Final-state correlation
effects (beyond LDA) 
are included only phenomenologically via the generalized
inner potential $V_{0,2}$.

v) The two-dimensional translational invariance of the surface is
obvious in the notations. The Green function and the matrix elements 
are diagonal with respect to ${\bf q}_\| = {\bf k}_\| + {\bf g}_\|$;
the same holds for spin projection $\sigma$. On the 
contrary, the problem does not separate with respect to layer and 
angular-momentum indices. However, the result allows to distinguish 
between the different partial contributions corresponding to hole 
propagation from $(i_\perp, L)$ to $({i_\perp}',L')$.

Apart from the advantage mentioned above, the main goal of the
present study has been to overcome the restriction to local 
potentials which is inherent in the original formulation of 
the one-step model. This is a necessary precondition to be able
to benefit from many-body theories which in general yield non-local 
self-energy corrections to the local LDA potential. Up to now these 
could not be used in the one-step model since they are incompatible 
with the KKR formalism for the initial state. If, on the contrary,
the Dyson equation for the Green function in a muffin-tin-orbitals 
representation is chosen as the starting point, an unproblematic 
connection to the results of many-body electronic-structure 
calculations is possible. Photoemission calculations based on this 
new concept are intended for the future.

\acknowledgements

This work was supported in part by the BMBF within the Verbundprojekt
``Elektronische Struktur und Photoemission von hochkorrelierten
intermetallischen seltenen Erdverbindungen'' 
(contract no.: 05605MPA0)
and in part by the Deutsche Forschungsgemeinschaft within the
Sonderforschungsbereich 290 (``Metallische d\"unne Filme: Struktur,
Magnetismus und elektronische Eigenschaften'').


\begin{references}

\bibitem[*]{ea}
corresponding author; address: Institut f\"ur Physik, 
Humboldt-Universit\"at zu Berlin, Invalidenstr. 110, D-10115 Berlin,
Germany; 
e-mail: potthoff@orion.physik.hu-berlin.de

\bibitem{FFW78}
B.~Feuerbacher, B.~Fitton, and R.~F. Willis, editors,
 {\em Photoemission and the Electronic Properties of Surfaces},
 Wiley, New York (1978).

\bibitem{CL78}
M.~Cardona and L.~Ley, editors,
 {\em Photoemission in Solids} volume~1,
 Springer, Berlin (1978).

\bibitem{Ing82}
J.~E. Inglesfield,
 Rep. Prog. Phys. {\bf 45}, 223 (1982).

\bibitem{CH84}
R.~Courths and S.~H\"ufner,
 Phys. Rep. {\bf 112}, 53 (1984).

\bibitem{Kev92}
S.~D. Kevan, editor,
 {\em Angle-Resolved Photoemission, Theory and Current Applications} 
 volume~74 of {\em Studies in Surface Science and Catalysis},
 Elsevier, Amsterdam (1992).

\bibitem{Dos83}
V.~Dose,
 Prog. Surf. Sci. {\bf 13}, 225 (1983).

\bibitem{Dos85}
V.~Dose,
 Surf. Sci. Rep. {\bf 5}, 337 (1985).

\bibitem{BT88}
G.~Borstel and G.~Th\"orner,
 Surf. Sci. Rep. {\bf 8}, 1 (1988).

\bibitem{Smi88}
N.~V. Smith,
 Rep. Prog. Phys. {\bf 51}, 1227 (1988).

\bibitem{Don94}
M.~Donath,
 Surf. Sci. Rep. {\bf 20}, 251 (1994).

\bibitem{Pen74}
J.~B. Pendry,
 {\em Low Energy Electron Diffraction},
 Academic Press, London (1974).

\bibitem{Pen76}
J.~B. Pendry,
 Surf. Sci. {\bf 57}, 679 (1976).

\bibitem{HPT80}
J.~F.~L. Hopkinson, J.~B. Pendry, and D.~J. Titterington,
 Comput. Phys. Commun. {\bf 19}, 69 (1980).

\bibitem{Pen80}
J.~B. Pendry,
 Phys. Rev. Lett. {\bf 45}, 1381 (1980).

\bibitem{MR80}
G.~Malmstr\"om and J.~Rundgren,
 Comput. Phys. Commun. {\bf 19}, 263 (1980).

\bibitem{GDG83}
B.~Ginatempo, P.~J. Durham, and B.~I. Gyorffy,
 J. Phys. C {\bf 1}, 6483 (1983).

\bibitem{TB84}
G.~Th\"orner and G.~Borstel,
 phys. stat. sol. (b) {\bf 126}, 617 (1984).

\bibitem{BTB85}
J.~Braun, G.~Th\"orner, and G.~Borstel,
 phys. stat. sol. (b) {\bf 130}, 643 (1985).

\bibitem{BTB87}
J.~Braun, G.~Th\"orner, and G.~Borstel,
 phys. stat. sol. (b) {\bf 144}, 609 (1987).

\bibitem{HTG+93}
S.~V. Halilov, E.~Tamura, H.~Gollisch, D.~Meinert, and R.~Feder,
 J. Phys.: Condens. Matter {\bf 5}, 3859 (1993).

\bibitem{TPF87}
E.~Tamura, W.~Piepke, and R.~Feder,
 Phys. Rev. Lett. {\bf 59}, 934 (1987).

\bibitem{SVH88}
B.~Schmiedeskamp, B.~Vogt, and U.~Heinzmann,
 Phys. Rev. Lett. {\bf 60}, 651 (1988).

\bibitem{SHHF94}
T.~Scheunemann, S.~V. Halilov, J.~Henk, and R.~Feder,
 Solid State Commun. {\bf 91}, 487 (1994).

\bibitem{HHSF94}
J.~Henk, S.~V. Halilov, T.~Scheunemann, and R.~Feder,
 Phys. Rev. B {\bf 50}, 8130 (1994).

\bibitem{GBB93}
M.~Gra\ss{}, J.~Braun, and G.~Borstel,
 Phys. Rev. B {\bf 47}, 15487 (1993).

\bibitem{GBB94a}
M.~Gra\ss{}, J.~Braun, and G.~Borstel,
 Surf. Sci. {\bf 46}, 107 (1994).

\bibitem{GBB94b}
M.~Gra\ss{}, J.~Braun, and G.~Borstel,
 Phys. Rev. B {\bf 50}, 14827 (1994).

\bibitem{HK64}
P.~Hohenberg and W.~Kohn,
 Phys. Rev. {\bf 136}, 864 (1964).

\bibitem{KS65}
W.~Kohn and L.~J. Sham,
 Phys. Rev. {\bf 140}, 1133 (1965).

\bibitem{SK66}
L.~J. Sham and W.~Kohn,
 Phys. Rev. {\bf 145}, 561 (1966).

\bibitem{GL76}
O.~Gunnarsson and B.~I. Lundqvist,
 Phys. Rev. B {\bf 13}, 4274 (1976).

\bibitem{HL71}
L.~Hedin and B.~I. Lundqvist,
 J. Phys. C {\bf 4}, 2064 (1971).

\bibitem{JG89}
R.~O. Jones and O.~Gunnarsson,
 Rev. Mod. Phys. {\bf 61}, 689 (1989).

\bibitem{SK95}
C.~M. Schneider and J.~Kirschner, 
 Critical Reviews in Solid State and Materials Science {\bf 20},
 179 (1995).

\bibitem{Bar92}
S.~D. Barrett, 
 Surf. Sci. Rep. {\bf 14}, 271 (1992).

\bibitem{Bra96}
J.~Braun, 
 Rep. Prog. Phys. {\bf 59}, 1267 (1996).

\bibitem{KV83}
W.~Kohn and P.~Vashishta,
 {\em {\rm In:} Theory of the Inhomogeneous Electron Gas},
 Ed. by S. Lundqvist and N. H. March, p. 79. Plenum, New York (1983).

\bibitem{AvB85}
C.-O. Almbladh and U.~von Barth,
 {\em {\rm In:} Density Functional Methods in Physics},
 Ed. by R. M. Dreizler and J. da Providencia, p. 209. Plenum, 
 New York (1985).

\bibitem{Bor85}
G.~Borstel,
 Appl. Phys. A {\bf 38}, 193 (1985).

\bibitem{Hed65}
L.~Hedin,
 Phys. Rev. {\bf 139}, 796 (1965).

\bibitem{Ary92}
F.~Aryasetiawan,
 Phys. Rev. B {\bf 46}, 13051 (1992).

\bibitem{RKP93}
M.~Rohlfing, P.~Kr\"uger, and J.~Pollmann,
 Phys. Rev. B {\bf 48}, 17791 (1993).

\bibitem{AG95}
F.~Aryasetiawan and O.~Gunnarsson,
 Phys. Rev. Lett. {\bf 74}, 3221 (1995).

\bibitem{NBDF89}
W.~Nolting, W.~Borgie{\l}, V.~Dose, and Th. Fauster,
 Phys. Rev. B {\bf 40}, 5015 (1989).

\bibitem{SOH90}
G.~Stollhoff, A.~M. Ole\'{s}, and V.~Heine,
 Phys. Rev. B {\bf 41}, 7028 (1990).

\bibitem{BN90}
W.~Borgie{\l} and W.~Nolting,
 Z. Phys. B {\bf 78}, 241 (1990).

\bibitem{BBN92}
J.~Braun, G.~Borstel, and W.~Nolting,
 Phys. Rev. B {\bf 46}, 3510 (1992).

\bibitem{SAS92}
M.~M. Steiner, R.~C. Albers, and L.~J. Sham,
 Phys. Rev. B {\bf 45}, 13272 (1992).

\bibitem{UIF94}
P.~Unger, J.~Igarashi, and P.~Fulde,
 Phys. Rev. B {\bf 50}, 10485 (1994).

\bibitem{NVF95}
W.~Nolting, A.~Vega, and Th. Fauster,
 Z. Phys. B {\bf 96}, 357 (1995).

\bibitem{VN96}
A.~Vega and W.~Nolting,
 phys. stat. sol. (b) {\bf 193}, 177 (1996).

\bibitem{Kor47}
J.~Korringa,
 Physica {\bf 6/7}, 392 (1947).

\bibitem{KR54}
W.~Kohn and N.~Rostocker,
 Phys. Rev. {\bf 94}, 1111 (1954).

\bibitem{Kam67}
K.~Kambe,
 Z. Naturforschg. {\bf 22a}, 322 (1967).

\bibitem{Kam68}
K.~Kambe,
 Z. Naturforschg. {\bf 23a}, 1280 (1968).

\bibitem{Kub57}
R.~Kubo,
 J. Phys. Soc. Japan {\bf 12}, 570 (1957).

\bibitem{Ada64}
I.~Adawi,
 Phys. Rev. {\bf 134}, A788 (1964).

\bibitem{Kel65}
L.~V. Keldysh,
 Sov. Phys. JETP {\bf 20}, 1018 (1965).

\bibitem{Mah70}
G.~D. Mahan,
 Phys. Rev. B {\bf 2}, 4334 (1970).

\bibitem{SA71}
W.~L. Schaich and N.~W. Ashcroft,
 Phys. Rev. B {\bf 3}, 2452 (1971).

\bibitem{CLRRSJ73}
C.~Caroli, D.~Lederer-Rozenblatt, B.~Roulet, and D.~Saint-James,
 Phys. Rev. B {\bf 8}, 4552 (1973).

\bibitem{FE74}
P.~J. Feibelman and D.~E. Eastman,
 Phys. Rev. B {\bf 10}, 4932 (1974).

\bibitem{HPM+95}
G.~Hilgers, M.~Potthoff, N.~M\"uller, U.~Heinzmann, L.~Haunert, 
 J.~Braun, and G.~Borstel,
 Phys. Rev. B {\bf 52}, 14859 (1995).

\bibitem{Eco90}
E.~N. Economou,
 {\em Green's Function in Quantum Physics},
 Springer, Berlin (1990).

\bibitem{pow88}
C.~J. Powell,
 J. Electron Spectrosc. Rel. Phen. {\bf 47}, 197 (1988).

\bibitem{PN95}
M.~Potthoff and W.~Nolting,
 Phys. Rev. B {\bf 52}, 15341 (1995).

\bibitem{PN96}
M.~Potthoff and W.~Nolting,
 J. Phys.: Condens. Matter {\bf 8}, 4937 (1996).

\bibitem{PN97c}
M.~Potthoff and W.~Nolting,
 Z. Phys. B (in press).

\bibitem{Mes79b}
A.~Messiah,
 {\em Quantenmechanik} volume~II,
 de Gruyter, Berlin (1979).

\bibitem{BGZ92}
W.~H. Butler, A.~Gonis, and X.-G. Zhang,
 Phys. Rev. B {\bf 45}, 11527 (1992).

\bibitem{Skr84}
H.~L. Skriver,
 {\em The LMTO Method, Muffin-Tin Orbitals and Electronic Structure} 
 volume~41 of {\em Spinger Series in Solid-State Sciences},
 Springer, Berlin (1984).

\bibitem{SS91}
S.~Y. Savrasov and D.~Y. Savrasov,
 unpublished (1991).

\end{references}
\end{document}